\documentclass[aps,prl,twocolumn,floatfix,showpacs]{revtex4}

\usepackage{graphics}
\usepackage{graphicx}
\usepackage{amssymb}
\usepackage{amsfonts}
\usepackage{amsmath}
\usepackage{bm} 

\begin{document}

\title{Fingering of Electron Droplets in Nonuniform Magnetic Fields} 
\author{Taylor L. Hughes} 
\author{A. D. Klironomos}
\author{Alan T. Dorsey}
\affiliation{Department of Physics, University of Florida, P.O. Box 118440, 
Gainesville, Florida 32611-8440}
\date{\today}

\begin{abstract}

A semiclassical analysis of a two-dimensional electron droplet in a high, 
nonuniform magnetic field predicts that the droplet will form ``fingered'' 
patterns upon increasing the number of electrons.  
We construct explicit examples of these patterns using methods 
first developed for the flow of two-dimensional viscous fluids.  
We complement our analytical results with Monte Carlo simulations of the
droplet wavefunction, and find that at the point where the 
semiclassical analysis predicts a cusp on the interface, the droplet
fissions---a type of ``quantum breakup'' phenomenon.

\end{abstract}


\maketitle

The growth of many physical systems is limited 
by diffusion---e.g., the diffusion of heat 
for solids growing into supercooled liquids
\cite{langer80}, the diffusion-limited aggregation (DLA)
of colloids \cite{witten83}, or the 
diffusion of magnetic flux in a superconductor \cite{frahm91}. 
The unstable growth of all of these systems \cite{langer80}  
produces beautiful patterns. 
An important paradigm for this class of pattern-forming systems is 
\textit{Laplacian growth} (LG)---diffusion-limited 
growth in the limit of long diffusion lengths
\cite{bensimon86}.  Agam \textit{et al.}\ \cite{agam02,wiegmann02}
recently discovered an intriguing connection between LG and
the growth of a two-dimensional (2D) electron droplet in a high, 
\textit{nonuniform}, magnetic field.  Their semiclassical 
analysis of the droplet wavefunction 
reveals that as the number of electrons
in the droplet is increased the droplet maintains a uniform
density while its boundary evolves according to the 
LG equations. The unstable growth leads to a ``fingered'' pattern!

We expand upon the work of Agam \textit{et al.}\ in 
several important ways.  First, we show that a simple 
magnetic field inhomogeneity \textit{outside} the growing droplet 
yields tractable analytical results for the droplet growth
which allow us to predict the onset and structure of interfacial 
singularities.  
Second, using an analogy between the probability density for the 
electrons and the Boltzmann weight for a fictitious 2D 
plasma in a background potential provided by   
the magnetic field inhomogeneity, we perform  
Monte Carlo (MC) simulations \cite{laughlin87} of the droplet growth. 
The droplet breaks apart beyond the point where the semiclassical 
analysis predicts a singularity, providing a quantum counterpart 
for droplet breakup in classical fluids \cite{shi94}. 
Finally, using the MC method we 
follow the evolution of the droplet interface in a random magnetic field 
and find an intricately fingered pattern. 

\begin{figure}
\includegraphics[width=6cm,clip]{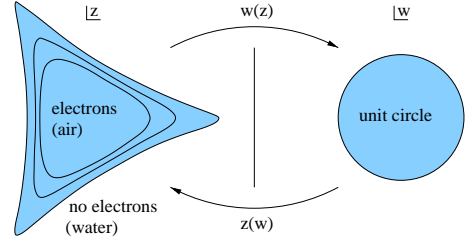}
\caption{
\label{fig1} 
Growth of a 2D electron droplet in an nonuniform magnetic field; 
on the left the interface is shown for
successively larger electron number.  In the Hele-Shaw analogy, 
the electron droplet corresponds to air which displaces water. 
}
\end{figure}

\textit{Hele-Shaw flow} provides a
simple example of LG.  
Two glass plates confine air and water, with 
the air injected into the center at a constant rate and the 
displaced water extracted at the edge, see Fig.~(\ref{fig1}). 
The air domain is at a constant pressure $p=0$ and  
the velocity of the water is determined
by Darcy's law, $\mathbf{v} = - \bm{\nabla} p$; since 
$\bm{\nabla}\cdot\mathbf{v}=0$, 
the pressure in the water is a harmonic
function of $z=x+iy$ with  a sink at infinity. On the interface
$v_n = -\partial_n p$, and in the idealized version of LG 
the surface tension is neglected and $p=0$ on the interface. 
Conformal mapping methods \cite{shraiman84,howison86} may be 
used to map the exterior of
the unit circle in an auxiliary $w$-plane onto the water 
domain in the $z$-plane at each 
instant of time $t$ using $z=z(w,t)$ (see Fig.~\ref{fig1}),
with $\mathrm{Re}(w \partial_w z\, \partial_t \bar{z}) = 1/2$ on the interface.

Next, consider a flat interface and produce a
sharp outward bump of the air into the water which locally compresses
the water isobars and produces a large pressure
gradient near the tip.   
Then $v_n = -\partial_n p$ implies that 
outward bumps grow more rapidly \cite{saffman58}
than the flat interface, 
and absent a stabilizing effect such as 
surface tension this bump develops into a singularity in finite time
\cite{shraiman84,howison86} for most initial conditions.

We next turn to the many-body wavefunction for 2D electrons 
in a perpendicular, 
\textit{nonuniform}, magnetic field.  If we ignore Coulomb interactions
and scale magnetic fields by the average field $B_0$, lengths 
by $l_b\equiv \sqrt{\hbar c/eB_0}$, and energies by 
$\hbar eB_0/2mc$, then 
the Hamiltonian is 
\begin{equation}
H = \sum_{j=1}^N\left[(-i\bm{\nabla}_j + {\bf A}_j)^2 
+ \frac{g}{2} \sigma_z B({\bf r}_j)\right].
\label{hamiltonian}
\end{equation} 
If $g=2$ then the many-body ground state wavefunction for Eq.~(\ref{hamiltonian})  
can be found \textit{exactly} even for
a nonuniform magnetic field \cite{aharonov79,bander90,girvin90};
for a spin-polarized, filled Landau level ($\nu=1$)  \cite{bander90,girvin90,agam02}
we have
\begin{equation} 
\Psi(z_1,\ldots,z_{N}) = \frac{1}{\sqrt{N! \tau_N} }\prod_{i<j} (z_i-z_j)
e^{\sum_j W(z_j)}, 
\label{many_body}
\end{equation} 
where $W(z) =-|z|^2/4 + V(z)$, with $V(z)$ a ``potential'' 
which solves $\nabla^2 V = - \delta B\equiv -[B(z) - B_0]/B_0$, 
and $ \tau_{N}$ is a normalization integral. 
Following Ref.~\cite{agam02} we assume that $\delta B=0$ in the 
region of the droplet; $V(z)$, however, is nonzero in 
this region and the electrons are influenced by distant field 
inhomogeneities---a manifestation of the Aharonov-Bohm effect 
\cite{wiegmann02}. In the region of the droplet $V(z)$ is an 
analytic function and has the expansion in terms of the 
harmonic moments $t_k$
\cite{dimensions}
\begin{equation} 
V(z) = \frac{1}{2} \mathrm{Re} \sum_{k=1}^\infty t_k z^k, 
\quad 
t_k = \frac{1}{\pi k} \int \delta B(z) z^{-k} d^2z. 
\label{field_moments}
\end{equation}

Although the wavefunction (\ref{many_body}) is exact 
for the noninteracting system, extracting physical 
observables such as the density is a formidable task. 
Agam \textit{et al.}\ \cite{agam02} perform a semiclassical 
analysis  of $\tau_N$ for $N\gg 1$ and find that
the integral is dominated by configurations in which the 
electrons are uniformly distributed in a domain 
of area $A=2\pi N$ with moments $t_k$ given by
Eq.~(\ref{field_moments}).  Further analysis establishes 
the connection with the LG problem and
provides us with powerful techniques \cite{howison86}
for determining the density distribution.  
 
The derivation of the wavefunction
(\ref{many_body}) requires two important 
assumptions---noninteracting electrons and 
$g=2$---which deserve a brief comment. 
First, for a \textit{uniform} magnetic field 
Laughlin's incompressible liquid 
state does an excellent job of capturing correlation effects 
due to the Coulomb interaction \cite{laughlin87}. 
However, this may not be correct when the magnetic field is 
\textit{nonuniform} as the Coulomb interaction
smoothes density fluctuations and will reduce the
fingering of the droplet, tending to make it circular. 
In fact, one can think of 
the Coulomb interaction as introducing an effective 
surface tension into the problem \cite{giovanazzi94}.  
In this case one can speculate that the fingering only 
occurs as a driven, nonequilibrium
phenomenon---the droplet must be constantly ``pumped'' by 
adding electrons to cause the droplet to finger and 
evolve \cite{wiegmann03}. 
The dynamical effects of Coulomb 
interactions for a \textit{classical} 
2D electron droplet were studied in Ref.~\cite{wexler99}; perhaps
this approach could be merged with the work of 
Agam \textit{et al.} in the semiclassical limit.  
The Coulomb interaction could also cause the droplet
edges to reconstruct \cite{chklovskii92}, which may 
also alter the fingering. 
Second,  $g\neq 2$ in semiconductor devices, and one might worry 
about the spin polarization of the droplet.  In uniform 
fields the Coulomb exchange interaction stabilizes the 
polarized $\nu=1$ state even for $g=0$ \cite{girvin98}, 
and it is likely that the same stability  occurs in nonuniform
fields, which would allow for a perturbative analysis in 
$g-2$ \cite{bander90}. It is unlikely
that $g\neq 2$ will produce any qualitative changes of the 
wavefunction (\ref{many_body}) and its analysis.  

As a concrete
example we will consider field inhomogeneities
which produce a monomial potential of the form 
\begin{equation}
V(z) = \frac{1}{2}\mathrm{Re}\,(t_{M+1} z^{M+1}),
\label{potential}
\end{equation}
with $M$ a positive integer.
This is the leading term of an interior multipole
expansion of $\delta B$.  
The beauty of
this choice for $V(z)$ is that the corresponding conformal map which 
describes the evolution of the interface in the semiclassical limit
has the simple form 
\begin{equation} 
z(w) = rw + u_{M} w^{-M},
\label{map}
\end{equation} 
where $r$ and $u_M$ are real functions of $N$.  Using either the 
moment method described above \cite{richardson72} or the 
Schwarz function method discussed in 
Refs.~\cite{howison86,mineev-weinstein00}, 
the parameters in Eqs.~(\ref{potential})
and (\ref{map}) can be related as 
\begin{equation} 
t_{M+1} = \frac{u_M}{(M+1) r^M}, \quad
t \equiv 2 N = r^2 - M u_M^2.
\end{equation}
Maps of the form (\ref{map}) have been 
studied \cite{howison86} for the Hele-Shaw problem, 
and many of the results can be directly applied to the electron 
droplet. For $M=1$, the interface is 
an ellipse which evolves smoothly, with constant 
eccentricity, for increasing $N$ \cite{difrancesco94}. 
The behavior for $M\ge 2$ is more interesting; 
the harmonic measure $w'(z)$ may have poles which coincide
with the interface at some critical value of the electron 
number $N^*$, resulting in singularities.   A detailed
analysis yields the following: 
(i) The interface develops
$M+1$ simultaneous cusps at 
\begin{equation} 
N^{*}=\frac{1}{2}(M-1)M^{-(M+1)/(M-1)}[(M+1)t_{M+1}]^{-2/(M-1)}.
\label{nstar} 
\end{equation} 
(ii) If $x^*=x(N^*)$ is the position of the cusp 
and $x(N)$ is the position of the finger for $N\alt N^*$, then 
\begin{equation} 
x^* - x(N) = \frac{2}{\sqrt{M-1}} \sqrt{N^* - N} + O(N^*-N),
\label{position}
\end{equation} 
so that the approach to the critical value is \textit{universal}, 
with a ``velocity'' 
$v_n \equiv dx/dN$ which diverges as $(N^*-N)^{-1/2}$.  
(iii) If we define the scaled variables
$X\equiv (x^* - x)/x^*$ and $Y\equiv y/x^*$, then the shape of
the cusp is universal and has the ``3/2'' form 
\cite{shraiman84,howison86} 
\begin{equation} 
Y(X) = \sqrt{\frac{2}{M}} \frac{M-1}{3} X^{3/2} + O(X^2). 
\label{shape}
\end{equation} 

These interfacial cusps arise from the neglect of
surface tension in the idealized LG problem.  For the 
Hele-Shaw problem the surface tension 
smoothes the cusp, while for the electron droplet one
expects that fluctuations beyond the semiclassical 
approximation will smooth the droplet \cite{agam02}.  
However, the mere existence of the singularity suggests that something
dramatic will happen to the droplet, and to investigate this we
need to go beyond the semiclassical level and obtain  
an essentially exact evaluation of the particle density
using the Monte Carlo method.

The probability density for
the electron droplet can be written as 
$|\Psi|^2 = \exp(-\beta U_\mathrm{cl})/N!\tau_N$,
with 
\begin{equation}
U_\mathrm{cl} = -\sum_{i<j}\ln |z_i-z_j|
+\sum_{i=1}^{N} \left[ \frac{1}{4}|z_i|^2 - V(z_i)\right],
\end{equation}
which is the Boltzmann weight for a classical 2D plasma in a 
background potential $-V(z)$, with a partition function $\tau_N$ 
and $\beta=2$.  The dominant contributions to
$\tau_N$ can be determined using the Metropolis
algorithm,  analogous to Laughlin's studies \cite{laughlin87}
of the quantum Hall effect in a \textit{uniform} 
field [i.e., $V(z)=0$].  An initially random distribution of 
$N$ particles is equilibrated
by moving each particle a distance of $2.0$ 
in a random direction, with moves accepted or rejected
according to the detailed balance condition; the average
acceptance rate for the moves is close to $0.5$.  
After an equilibration period of about
$10^3$ moves per particle we compile a histogram of 
particle positions in bins of size $0.3\times 0.3$
for about $4\times 10^4$ moves per particle, and then 
calculate the average density in each bin.  
The electrons were confined by the walls of an 
impenetrable ``simulation box'' whose dimensions 
were chosen to be large compared to the linear
dimension of the droplet.
We accurately reproduce known results 
\cite{laughlin87} for the droplet 
size, density, and energy for a uniform field. 
For a nonuniform field the density forms a 
complicated pattern, but it 
still has a uniform value $\rho_\mathrm{ave}$
well away from the interface. 
We determine the interface by finding the locus of points
whose density is between, say, $0.3\rho_\mathrm{ave}$ and
$0.4 \rho_\mathrm{ave}$ (other definitions give similar results).

\begin{figure}
\includegraphics[width=7cm,clip]{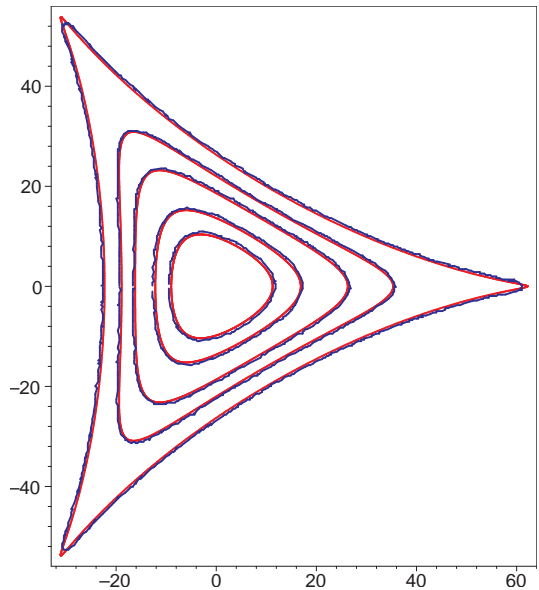}
\caption{\label{fig2}
Comparison of semiclassical (red) and Monte Carlo 
results (blue) for the droplet interface with $M=2$; 
starting from the innermost curve,
$N=50$, 100, 200, 300, and 494. Three cusps are predicted to form at
$N^*=500$.  
}
\end{figure}

\begin{figure}
\includegraphics[width=6.7cm,clip]{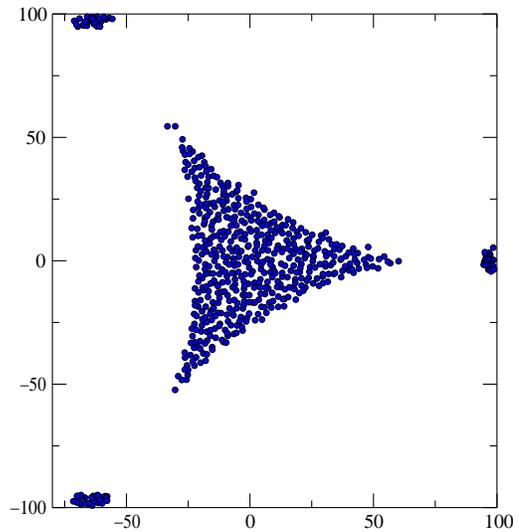}
\caption{\label{fig3} Snapshot of particle positions 
in a simulation for $M=2$, $N^*=500$, and $N=600$.  
Eighty nine
electrons have split from the central droplet and 
are organized into three smaller droplets 
at the boundary of the simulation box. 
}
\end{figure}

\begin{figure}
\includegraphics[width=7cm,clip]{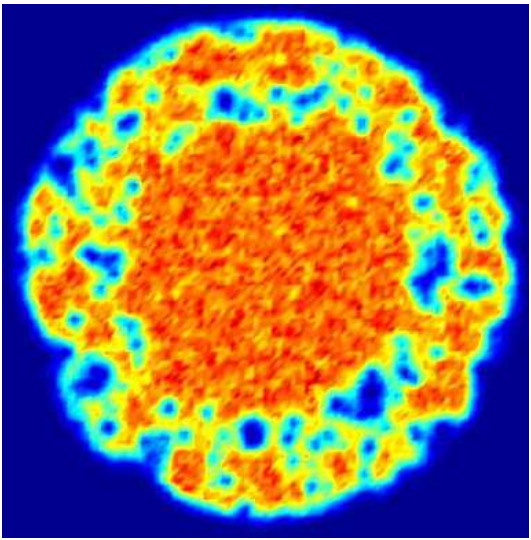}
\caption{\label{fig4} 
Average electron density for 280 electrons in a random array of 400 solenoids; 
blue is low density and red is high density.  In this simulation 50 
``attractive'' solenoids were placed far from the droplet, with 350 
``repulsive'' solenoids placed closer to the droplet.  
The total solenoid flux is zero.
} 
\end{figure}

We have performed numerous MC simulations for ordered and
random magnetic field configurations, with the 
results shown in Figs.~(\ref{fig2})--(\ref{fig4}).  Let's start with 
the potential $V(z)= (t_3/2) z^3$ for $M=2$, which may
be thought of as the leading term in a multipole expansion for an
arrangement of six thin solenoids with alternating flux $\pm \Phi$ placed
on the vertices of a hexagon a large distance $R$ from the 
center of the droplet, so that $t_3=4\Phi/R^3$. 
The semiclassical analysis predicts that three 
cusps will appear at the 
critical electron number $N^* = 1/(144 t_3^2)$, and
we have chosen $N^*=500$ for these simulations. 
The excellent agreement between the semiclassical
and MC results can be seen in Fig.~(\ref{fig2});
for $N=494$ the nascent MC cusp is slightly rounded 
compared to the semiclassical prediction due to  the 
discrete nature of the particles. 
We find equally impressive agreement with the position of
one of the tips, Eq.~(\ref{position}), and
the shape of the cusp, Eq.~(\ref{shape}). 
When $N>N^*$ the electrons in excess of $N^*$ 
move toward the edges of the simulation box 
as shown in Fig.~(\ref{fig3}), 
so that the central droplet fissions. 
In the 2D plasma language, the 
incompressible fluid of particles saturates the 
potential minimum and additional particles
which are ``poured'' into the potential 
``spill out'' and accumulate at the edge of the simulation box.  
Finally, Fig.~(\ref{fig4}) shows a simulation for a 
random distribution of thin solenoids, with a  
highly ramified pattern for the averaged electron density.
Determining whether such a pattern is fractal 
requires a study of the scaling law 
$N\sim R^{D_f}$, with $R$ the radius of gyration and 
$D_f$ the fractal dimension (with $D_f=2$ for compact and 
$D_f<2$ for fractal patterns).  An unambiguous determination of
$D_f$ requires simulations over many decades in $N$; 
this is currently beyond our computational resources and we 
leave its resolution as an interesting open problem. 
               
In summary, we have constructed several explicit examples of magnetic 
field inhomogeneities which cause a noninteracting 2D electron droplet 
to finger.  The semiclassical results (idealized LG) are in excellent 
agreement with our Monte Carlo simulations of the droplet wavefunction.  
An important implication of this result is that if there is an 
effective ``surface tension'' in the noninteracting 
electron droplet problem it must be extremely small, and the 
regularization of the cusps appears to be controlled by the discreteness of
the particles. 
Beyond the semiclassical singularity we find that the droplet fissions. 
Recent advances in imaging techniques for the 2D electron 
gas \cite{finkelstein00} may make it possible to 
observe these fingered structures; an easily tuned field inhomogeneity
could be provided by placing the electron gas in close
proximity to a type-II superconductor in the vortex
state \cite{bander90}.  Our results also raise the interesting
question of whether equilibrium Monte Carlo methods using a
random $V(z)$ can be used to efficiently simulate nonequilibrium 
growth processes such as DLA \cite{witten83}; 
in other words, what can the quantum Hall effect tell us about DLA?

We would like to thank S. Girvin, R. Goldstein, F. Klironomos,
and L. Radzihovsky  for their helpful comments, and 
P. Wiegmann for inspirational correspondence.
TLH was supported by NSF REU grant DMR-0139579, and ADK and 
ATD were supported by NSF DMR-9978547.


\begin{thebibliography}{99}
\bibitem{dimensions}
Our definition of $t_k$ is twice that of Ref.~\cite{agam02}
due to a different choice of units.
\end{thebibliography}
\end{document}